\documentclass[amsmath,amssymb,preprint, aps]{revtex4}
\usepackage{amsmath}
\usepackage{amssymb, chemarr}
\usepackage{graphicx}
\pdfoutput=1

\begin{document}
\newcommand{{\calR}}{{\cal{R}}}

\title[Fan-out in Gene Regulatory Networks]{Fan-out in Gene Regulatory Networks}

\author{Kyung Hyuk Kim$^*$ and Herbert M. Sauro\\
		Department of Bioengineering,
	   University of Washington,\\
	   William H. Foege Building,
	   Box 355061, 
	   Seattle, WA 98195-5061, U.S.A.
	   \vspace{2cm}
	   }

\begin{abstract}
 In synthetic biology, gene regulatory circuits are often constructed by combining smaller circuit components.  Connections between components are achieved by transcription factors acting on promoters.  If the individual components behave as true modules and certain module interface conditions are satisfied, the function of the composite circuits can in principle be predicted. 
In this paper, we investigate one of the interface conditions: fan-out.   We quantify the fan-out, a concept widely used in electric engineering, to indicate the maximum number of the downstream inputs that an upstream output transcription factor can regulate.  We show that the fan-out is closely related to retroactivity studied by Del Vecchio, et al.  We propose an efficient operational method for measuring the fan-out that can be applied to various types of module interfaces.  We also show that the fan-out can be enhanced by self-inhibitory regulation on the output. We discuss the potential role of the inhibitory regulations found in gene regulatory networks and protein signal pathways. 
The proposed estimation method for fanout not only provides an experimentally efficient way for quantifying the level of modularity in gene regulatory circuits but also helps characterize and design module interfaces, enabling the modular construction of gene circuits.
\vspace{2cm}
\\
$^*$ Corresponding Author: kkim@uw.edu
\\
Running title: Fan-out in Gene Regulatory Networks
\\
Character count: 53631
\end{abstract}


\maketitle
\clearpage

\section*{Background}
Engineering relies on modular composition that is the ability to combine functional units with the knowledge that the intrinsic properties of each module is unaffected to a large degree by the composition. In biology we are less clear on the notion of a modular component, or at least biology has multiple definitions depending on context. Here we will define a module as a self-contained functional unit whose intrinsic properties are independent of the surrounding milieu. This definition is similar to that used in engineering. For example, the intrinsic properties of a TTL NAND gate~\cite{TTLCookbook} is unaffected (within certain design constraints) when connected to other TTL logic gates. That is, a NAND gate remains a NAND gate no matter what it is connected to. This property allows engineers to design, predict and fabricate complex circuits at very low cost. The question whether such self-contained and functionally independent modules exist at the biological cellular network level is still an ongoing research problem \cite{Alon2007}. In this paper we will be concerned with the design of modular synthetic components~\cite{Purnick2009, Lucks2008, Keasling2008, Voigt2006, Sprinzak2005, Endy2005} and avoid the question of modularity in natural complex systems.

In the most abstract sense we can define a module as follows. Given a functional unit $M$ with input $I$ and output $O$, we can define a relation between the input and output as $O = M(I)$. Given two functional units, $M_1$ and $M_2$, where the output of $M_1$ serves as the input to $M_2$, then $M_1$ and $M_2$ are defined as modules if the relation, $O_2 = M_2 (M_1 (I_1))$ is true. This simply means that in connecting $M_1$ and $M_2$ together, $M_2$ has no effect on the functional characteristics of $M_1$ and vice versa.  

Predictable composition is of particular interest to the synthetic biology community where gene circuits are ``wired" together via transcription factors (TFs) and corresponding promoters (Figure~1). This mode of wiring makes the physical construction of relatively complex networks possible \cite{Hasty2002,Stelling2004}. However the general question of whether making a connection between two genetic units results in a functional whole remains.  In particular a number of issues present themselves that include independence from the surrounding milieu (also called orthogonality \cite{Lucks2008}): domain matching and impedance bridging. The former describes the situation where the operating concentration range of an output transcription factor matches the range of the input target.  Impedance bridging,  which is the main  topic of this paper,  is concerned with how much a downstream target circuit can affect the functional properties of an upstream unit.  It is not to be confused with impedance matching which is related to the maximum power transfer between two circuits. There has recently been interest in defining impedance bridging in genetic and protein circuits \cite{Sauro2004b,Sauro2007} and a related quantity called retroactivity was introduced by Saez-Rodriguez et al.~\cite{Saez2005} and Del Vecchio et al.~\cite{Vecchio2008} to describe the effect of one module on another.

In electric engineering  there exist guidelines and published constraints on how many electrical modules can be driven from a source. For example, one rule of thumb for analog circuits suggests that the impedance at the input should be ten times the impedance at the driving circuit. In digital circuits, such as TTL circuits, manufacturers will quote the fan-out and fan-in for a given electrical module.  The fan-out indicates how many downstream logic gates can be connected to a given output. Exceeding these limits will potentially cause signal distortion in analog circuits and circuit failure in digital circuits. 

We envision the development of similar criteria for connecting two biological modules together in synthetic biology, and here introduce the notion of fan-out for a genetic circuit. We define the fan-out of a genetic circuit as the maximum number of downstream promoters that can be driven from an upstream circuit signal without significant time-delay or signal attenuation.  We will show that the fan-out can be estimated by using the autocorrelation \cite{Anishchenko, Simpson2003, Rosenfeld2005, Austin2006, Weinberger2008, Dunlop2008} of gene expression noise  \cite{Rao2002, Raser2004, Kaern2005, Shahrezaei2008b}.  During the estimation procedure the system's retroactivity can also be measured.  This fan-out/retroactivity estimation can be applied under quite general module interface conditions.  Although our analysis is focused on genetic networks, the principles apply equally to signal transduction networks.

\section*{Results and Discussion}
 \subsection*{Module interface process}\noindent
When two synthetic gene circuits are connected, transcription factors are used to connect them.  The reaction processes involving the transcription factors such as transcription, translation, degradation, and downstream-module promoter regulation, will be called \emph{module interface processes} (MIPs).  For example, consider the repressilator \cite{Elowitz2000}.  Let us choose the TetR repressor, one of the genes comprising the oscillator, as an output of the oscillator module (Fig.~1).  When a downstream module has \emph{tetR}-operons, the MIP includes \emph{tetR}-transcription, translation, and TetR binding/unbinding to its specific operons located in the downstream module.

\subsection*{Retroactivity and mapping between a module interface process and an RC-circuit}\noindent
We will investigate a MIP  by  mapping it to a simple electric circuit composed of a resistor and a capacitor connected in series (RC circuit).  This mapping  becomes significantly helpful for understanding retroactivity \cite{Vecchio2008} and quantifying fan-out.  

\subsubsection*{Isolated case}\noindent
When an upstream output does not regulate any downstream promoter, we can model the corresponding MIP as a simple TF translation-degradation process (see Fig.~2A and B).  The concentration of the TF, denoted by $X$, changes in time by following the equation
\begin{equation}
\frac{dX}{dt} = \alpha (t) - \gamma X,
\label{eqn:TF-isolation}
\end{equation}
with $\alpha(t)$ the translation rate and $\gamma$ the degradation rate constant. 
We will show how this process can be related to an RC circuit, where a resistor and capacitor are connected in series and driven by an input voltage source $V_{in}$ (Fig.~2C).  The total voltage drop across both the resistor and capacitor is equal to the driven voltage: $V_{in} = RI + V_{out}$,
where $I$ denotes the current flowing through the resistor, and $V_{out}$ the voltage drop across the capacitor.  The current  is equal to the rate of charge accumulation ($Q$) in the capacitor: $I=dQ/dt$, where the small increment $dQ$ causes the change in $V_{out}$ in proportion to $dQ$: $dQ=C dV_{out}$, with $C$ a proportionality constant called capacitance.  Thus, the current $I$ can be expressed as $C dV_{out}/dt$.  By substituting this into $V_{in} = RI + V_{out}$ and dividing the resultant equation by $RC$, we obtain
\begin{equation}
\frac{dV_{out}}{dt}= \frac{V_{in}}{RC} - \frac{V_{out}}{RC},
\label{eqn:RC-isolation}
\end{equation}
where $RC$ is known as the response time $\tau_0$ of the RC-circuit \cite{note-retroact}. 
By comparing Eqs.~\eqref{eqn:TF-isolation} and \eqref{eqn:RC-isolation}, we obtain the following correspondence: $X = V_{out}$, $\alpha = V_{in}/RC$, and $\gamma = 1/RC$,
and the response time is expressed as 
\begin{equation}
\tau_0 = RC = \frac{1}{\gamma}.
\label{eqn:tau0}
\end{equation}
Thus, the TF-translation-degradation process (Fig.~2B) can be directly mapped to the RC-circuit (Fig.~2C).

\subsubsection*{Connected case}\noindent
Now we consider the case where two modules are connected (see Fig.~3A).  We will review retroactivity studied by Del Vecchio et al.~\cite{Vecchio2008}.  The retroactivity was defined as the slow-down in interfacial dynamics in response to interactions with downstream components. The retroactivity described a number between zero and one with one being the least desirable, i.e. the interfacial dynamics are affected most. In their analysis, they assumed that the binding-unbinding process of the TF is fast enough that the process can be approximated to be in the quasi-steady state ($k_{on}X + k_{off} \gg \gamma$; cf. \cite{Kepler2001, Rao2003, Simpson2004}).  They also assumed that the lifetime of the bound TF is  much longer than that of the unbound TFs.  

Specifically, they showed that the free TF concentration $X$ changes in time by the following equation \cite{Vecchio2008}
\begin{equation}
\frac{dX}{dt} = (1-\calR(X))(\alpha - \gamma X),
\label{eqn:TF-connected}
\end{equation}
where $\calR (X)$ is the \emph{retroactivity}, given by $ \left[1+\left(  1+\frac{X}{K_d} \right)^2 \frac{K_d}{P_T} \right]^{-1}$, with $K_d$  the dissociation constant for the TF with respect to the promoter, and $P_T$ the total number of the promoters.  They showed that $\calR$  is always less than 1 and non-negative.  The extra factor $1-\calR$ appears when compared with the isolated case, resulting in the slow-down of the dynamics.  More precisely, the slow-down is due to the decrease in the factor placed in front of $X$ in Eq.~\eqref{eqn:TF-connected}: $\gamma (1-\calR)$, which is related to the apparent response time: 
\begin{equation}
\tau_a \equiv \frac{1}{(1-\calR)\gamma}.
\label{eqn:taua}
\end{equation}

We will consider the MIP shown in Figs.~3A and B under the same assumptions as given by Del Vecchio et al \cite{Vecchio2008}.  To understand the retroactivity by using an RC-circuit analogy, consider a circuit as shown in Fig.~3C.  The total capacitance becomes the sum of the two capacitances: $C_T = C+C'$.  Thus, the response time becomes $RC_T$: $\tau = RC_T$.
The change in the output voltage is governed by the same equation as in the isolated case except the capacitance $C$ is replaced to $C_T$:
\begin{equation}
\frac{dV_{out}}{dt} = \frac{V_{in}}{RC_T} - \frac{V_{out}}{RC_T} = \left[  1-\frac{C'}{C+C'}  \right] \left[   \frac{V_{in}}{RC} - \frac{V_{out}}{RC} \right].
\label{eqn:RC-connected}
\end{equation}
By comparing Eqs.~\eqref{eqn:TF-connected} and \eqref{eqn:RC-connected}, we find that the retroactivity is given by the relative ratio of the new capacitance: 
\begin{equation}
\calR = \frac{C'}{C+C'},
\label{eqn:retro-c}
\end{equation}and that the response time $\tau$ corresponds to  $\tau_a$ (Eq.~\eqref{eqn:taua}):
\begin{equation}
\tau = RC_T = \frac{1}{(1-\calR)\gamma}.
\label{eqn:tau-rct}
\end{equation}

We have shown that connecting downstream promoters in the MIP is equivalent to connecting extra capacitors in parallel with an existing one in the RC-circuit. Due to these extra capacitors, the circuit takes a longer time to fully charge all the capacitors, resulting in the slow-down in the circuit response time.  
Biologically, the bound promoters act as a reservoir of potentially free TFs: Whenever there is a change in the number of the free TFs, the reservoir quickly buffers the change in the number of free TFs~\cite{Buchler2009}.  Such buffering causes transient dynamics at the interface to slow down.

\subsection*{Response time vs. number of promoters}\noindent
We now investigate how the response time $\tau$ is related to the total number of promoters $P_T$.  The response time is shown to increase with $P_T$ (see  the Methods section) as
\begin{equation}
\tau_{P_T} = R(C+P_T C_1),
\label{eqn:response-time}
\end{equation}
where $C_1$ is a proportionality constant satisfying $C/C_1 = K_d \left(  1+ X/K_d  \right)^2$.
The above equation~\eqref{eqn:response-time} can be viewed as each individual promoter contributing an extra capacitance $C_1$ to the total capacitance (Fig.~3C): $C_T = C+P_T C_1$.
The capacitance $C_1$ of each extra capacitor is related to a unit load onto the upstream output dynamics from a single downstream promoter.  This is an interesting result and becomes useful for proposing an experimental method for estimating the fan-out.

The linear relationship between the extra capacitance and $P_T$ (see Fig.~3C box) does not come from any linearization approximation, but from the fact that each downstream promoter affects the upstream as an independent effector (reservoir or sequestrator), although the sequestration itself is represented by a nonlinear reaction.

\subsection*{Gene circuit fan-out}\noindent
We will define a gene circuit fan-out by the maximum number of promoters in a downstream module  that the output (transcription factor) of an upstream module can regulate without altering the output dynamics significantly.  To exemplify how much the upstream module can be affected, we consider a repressilator \cite{Elowitz2000} as a module and its Tet repressors as a module output (Fig.~1A). When the output regulates \emph{tetR} promoters located in a downstream module, the oscillation amplitude of the \emph{tetR} expression level can be significantly changed, e.g., 40\% decrease when  the number of the promoter ($P_T$) is changed from $0$ to $100$ (Fig.~1B and C).  Our interest is here to quantify the maximum number of the promoters (fan-out) that  the upstream module can tolerate.

Let us quantify the fan-out by considering again the simple MIP shown in Fig.~3A.  We consider a frequency response between the input and output voltage,  $V_{in}$ and  $V_{out}$, respectively (Fig.~4A).   In the RC circuit, the capacitor acts as a low pass filter: The gain of the signal (the ratio of the oscillation amplitude of $V_{out}$ to that of $V_{in}$) is at the maximum level for low frequencies and drops significantly when the circuit no longer responds as fast as the input signal changes (Fig.~4B).  The frequency when this happens is called \emph{cut-off frequency} ($\omega_c$) (Fig.~4B) and corresponds to the inverse of the response time: $1/RC_T$ \cite{Nilsson2008}.

If  the upstream module functions as a synthetic oscillator, there will be a practical upper limit ($\omega_{max}$) in the oscillator's frequency (e.g., for the repressilator, $\omega_{max}$ can be the inverse of the repressor lifetime $= \log(2)/10$ min$^{-1}$  $\sim 4$ hour$^{-1}$ \cite{Elowitz2000}). If $\omega_{max}$ is smaller than the cut-off frequency $\omega_c$, the oscillator output will operate in a predictable manner and the output signal will be passed downstream without any significant signal loss.  As the number of the downstream promoters increases, the total capacitance increases as shown in Eq.~\eqref{eqn:response-time} and the cut-off frequency ($\omega_c = 1/RC_T$) decreases.   For the cut-off frequency to be larger than the maximum operational frequency $\omega_{max}$, the total number of the promoter must be smaller than a certain value, which will be called  the \emph{fan-out}.  The fan-out denoted by $F_{\omega_{max}}$ is obtained where $\omega_c$ equals $\omega_{max}$, i.e.,  $\omega_c = [ R(C+P_T C_1)]^{-1} = \omega_{max}$: 
\begin{equation}
F_{\omega_{max}} = \frac{C}{C_1} \left[ \frac{1/\tau_0}{\omega_{max}} -1  \right].
\label{eqn:fan-out}
\end{equation}

In the fan-out equation~\eqref{eqn:fan-out}, there are two unknown parameters: $C/C_1$, and $\tau_0$.   These can be experimentally estimated by performing two independent experiments with and without any downstream module.  In each experiment we measure the corresponding response time: $\tau_0$ and $\tau_{P_T}$ (by using gene expression noise as will be presented later in the Results section).  Thus, one of the unknowns $\tau_0$ can be estimated.  How can we estimate the other unknown $C/C_1$ from $\tau_{P_T}$?  If we know a priori  the copy number of the promoters $P_T$, we can obtain the value of $C/C_1$ from Eq.~\eqref{eqn:response-time}.    If the promoters are placed on plasmids, the copy number of the plasmids can be estimated depending on what type of origin of replication is used, and thus the copy number of the promoters $P_T$ can be known.   
By calculating $\tau_0/RC_1$, we can estimate the other unknown, $C/C_1$.

\subsection*{Gene circuit fan-out in more general interfaces}\noindent
Up to now we have considered a simple MIP without feedback and where the degradation rate is assumed to be first-order.   Here we will consider the more general case and show that we can use the same or similar fan-out function as Eq.~\eqref{eqn:fan-out}.  

\subsubsection*{Oligomer under directed degradation and self-regulation }\noindent
Consider that a TF, composed of $n$ monomers, is tagged for degradation and that its transcription is self-regulated as shown in Fig.~5A.  We can obtain the fan-out expressed by the same equation~\eqref{eqn:fan-out}, where $\tau_0$ is the time constant in the isolated case, given by the difference between the elasticities \cite{Fell1992}: $1/\tau_0 = \epsilon_2 - \epsilon_1$, where $\epsilon_1 \equiv \partial v_1(x,\alpha) / \partial x$ and $\epsilon_2 \equiv \partial v_2(x)/ \partial x$ (refer to the Methods section).   This means that the fan-out can be estimated exactly in the same way as in the monomer case as shown in Fig.~3A.    All the above results apply for the case that the \emph{intermediate} reaction steps of the oligomerization and directed degradation are taken into account (refer to the SI). 

\subsubsection*{Multiple promoters having different affinities}\noindent 
 Consider the case that two different types of TF-specific promoter plasmids, having different affinities for the TF and different strength of the origin of replication.  It is shown that the MIP can be mapped to an RC-circuit having two different capacitances connected in parallel to $C$ as shown in Fig.~5B (see the Methods section).  The fan-out of each promoter is shown to satisfy the following functional relationship between $F_1$ and $F_2$ (refer to the Methods section): 
\begin{equation}
1/\omega_{max} = \tau_0(1+F_1 \frac{C_1}{C} + F_2 \frac{C_2}{C}),
\label{eqn:fan-out-gen}
\end{equation}
where $F_{i}$ is the fan-out for promoter plasmids of the $i$-th kind, and $C_i$ denotes the corresponding capacitance per plasmid.  If there are $N$ different kinds of promoter plasmids, we need to sum up all the capacitances in the above equation.   We note that the fan-out is not a single number but is given by a functional relationship between $F_i$'s: We need to balance the number of plasmids of different kinds depending on its unit load on the retroactivity, i.e., $C_i/C$.   

To obtain the fan-out function, we need to find three unknown parameters: $\tau_0$, $C_1/C$, and  $C_2/C$.  $\tau_0$ can be estimated in the isolated case.  $C_i/C$ can be estimated in the case that only the $i$-th kind of promoter plasmids exists (under the assumption that the strength of each origin of replication is already known).   These three independent experiments will suffice for estimating all the unknown parameters and proposing the fan-out function Eq.~\eqref{eqn:fan-out-gen}.

\subsubsection*{Multiple operators}\noindent
Consider the case that the promoter region includes multiple operators specific to an output TF (e.g., $O_1$, $O_2$, and $O_3$) having different affinities (Fig.5C).  Regardless the number of the operators, the same fan-out function as Eq.~\eqref{eqn:fan-out} is obtained (refer to the SI). 

\subsubsection*{Two output signals}\noindent
When two output TFs ($X$ and $Z$) regulate a downstream promoter independently, i.e., if there is no overlap between the operator regions and somehow $X$ does not interfere with the operator region of $Z$ and vice versa, the fan-out corresponding to each output TF can be obtained.

We have shown that the fan-out function like Eqs.~\eqref{eqn:fan-out} and \eqref{eqn:fan-out-gen} can be obtained for each of individual cases given above.  Furthermore, for all the combinations of these individual cases the fan-out functions apply.

\subsection*{Design scheme for fan-out enhancement}\noindent
How can we increase the fan-out?  Based on the fan-out equations~\eqref{eqn:fan-out}, there are two ways: increasing $C/C_1$ or $1/\tau_0$.  The way to increase the latter is to apply a negative feedback on the translation of $X$ (making $\epsilon_1$ negative for the case shown in Fig.~5A, where $1/\tau_0 = \epsilon_2 - \epsilon_1$) and  a positive feed-forward on the degradation rate (increasing $\epsilon_2$).   Since the enhanced degradation and negative feedback decrease the concentration level of $X$, to prevent this, it is desirable to amplify the translation rate (which makes $\epsilon_1$ more negative).  This is exactly the same mechanism proposed by Del Vecchio et al. \cite{Vecchio2008} to reduce retroactivity; when the retroactivity is small, the upstream output dynamics does not slow down significantly by connecting the output to the downstream module, meaning that the load from downstream to the upstream is small enough that many replicates of the load can be applied to the upstream without slowing down the output dynamics significantly.

One of the mechanisms, inhibitory auto-regulation, is frequently found in \emph{Escherichia coli} transcription factors regulating a set of operons, e.g., for  amino-acid biosynthesis where a single TF may control multiple targets, likewise for flagella formation \cite{Shen2002}.  Such motifs are called sing-input-module motifs \cite{Shen2002}.  

The concept of fan-out is not limited to gene regulatory circuits.  In principle, as long as the same class of interface processes are found regardless of the type of biological systems, the fan-out and retroactivity concepts can be applied  \cite{Sauro2008, Vecchio2008}.  For example, in the eukaryotic MAPK pathway, doubly phosphorylated MAPK can activate a number of downstream proteins and transcription factors in the nucleus.  This MAPK regulation can be described by the module interface process similar to the one shown in Fig.~5B (in this case, many promoter plasmids instead of the two).   In the MAPK pathway, there is a negative feedback from MAPK to the phosphorylation of MAPKKK~\cite{Sauro2004b,Sauro2007}.  The negative feedback increases the fan-out of the MAPK module thereby permitting MAPK to effectively regulate multiple targets and multiple homologous binding sites.

\subsection*{How to measure the time constant $\tau$}\noindent
It is  known that  transcription factors show significant stochastic fluctuations \cite{Arkin1998, Elowitz2002, Ozbudak2002,Elf2007,Rosenfeld2005, Austin2006,  Weinberger2008, Dunlop2008} (for review articles, \cite{Rao2002, Raser2004, Kaern2005, Shahrezaei2008b}).  Their correlation times have  been measured by obtaining autocorrelations by \emph{in vivo} time-lapse microscopy \cite{Rosenfeld2005, Austin2006, Weinberger2008, Dunlop2008}.  Recent numerical studies show that the correlation time is approximately equal to the response time of the deterministic case \cite{Kim2009} and that it changes as a result of connecting two genetic systems \cite{Kim2009, Jayanti2009}. Therefore,
by using the change in the correlation time, we can estimate the fan-out by using Eq.~\eqref{eqn:fan-out} as well as the retroactivity by using $\calR = \frac{\tau_{P_T}- \tau_0}{\tau_{P_T}}$ (obtained from 
Eq.~\eqref{eqn:retro-c} by using Eqs.~\eqref{eqn:tau0} and \eqref{eqn:tau-rct}).

\subsection*{Example: Fan-out/retroactivity estimation}
\noindent 
Let us consider the simple MIP shown in Fig.~3A as a model for TFs in  \emph{E.~coli}.  We consider that the average copy number of plasmids containing the specific promoters ranges from $1$ to $100$.  If we set the volume of  \emph{E.~coli} roughly equal to $1\mu\mbox{m}^3$, a copy number of one corresponds to 1~nM.  As a result we will henceforth interchange the unit of nM  with that of copy number. A simulation using the standard Gillespie method \cite{Gillespie1977} was performed and the observed autocorrelation was fitted to an exponential function: $G(\Delta t)=A \exp(-\Delta t/\tau)$ with $\tau$ a correlation time (we conducted a linear fit in the log-scale in the $y$-axis and the normal scale in the $x$-axis) and $1/\tau$  obtained from the fitted slope. 

For experimentally reasonable parameter values, i.e., $\alpha=20$ nM hour$^{-1}$, $\gamma=2$ hour$^{-1}$, $k_{on}= 10$ nM$^{-1}$hour$^{-1}$, and $k_{off}=10$ hour$^{-1}$, we performed the stochastic simulations with and without any downstream-module promoter ($P_T=100$ and $0$).  The concentration levels of the total TF was recorded 100 times over 2 hours, the autocorrelation of this signal fitted to an exponential function, and the response time measured \cite{Kim2009}.  We obtained the error bar of the time constant from 10 independent replicates of the autocorrelation.  

When the translation rate was set to 20 nM hour$^{-1}$,  we obtained $\tau_0=0.52 \pm 0.06$  hour and $\tau_{100}=0.9 \pm 0.1$ hour.  We obtained $C/C_1 = 140 \pm 20$, by using 
\[
\frac{C}{C_1} = P_T\frac{RC}{RC_T-RC}= \left. P_T \frac{\tau_0}{\tau_{P_T} - \tau_0} \right|_{P_T=100},
\]
where we used $C_T-C = P_T C_1$.   From Eq.~\eqref{eqn:fan-out},  we obtained the fan-out function for this MIP:
\[
F_{\omega_{max}} = 140[\pm20] \left( \frac{1/0.9[\pm0.1]}{\omega_{max}}-1\right).
\]
If the upstream module is a synthetic oscillator with a maximum operating frequency $\omega_{max}=1$ hour$^{-1}$, the fan-out becomes $F =130 \pm 20$.  This means that we can use promoter plasmids with low, medium, and high copy numbers without affecting the TF dynamics, if a single TF-specific operator site resides on a plasmid. The retroactivity can also be estimated from the measured values of $\tau_0$ and $\tau_{100}$: $\calR = 0.4 \pm 0.1$.

If we reduce the translation rate by half (now, $\alpha=10$ nM hour$^{-1}$) the free TF concentration decreases by half.  As the concentration decreases, the retroactivity increases \cite{Vecchio2008, Kim2009} and the fan-out decreases.  We estimated $\tau_0=0.52 \pm 0.07$ hour and $\tau_{100}=1.75 \pm 0.04$ hour.   For the same $\omega_{max}=1$ hour$^{-1}$ we obtained the fan-out: $F= 40 \pm 1$. This would mean that we could safely use only low copy number plasmids. The retroactivity is estimated to be $0.70 \pm0.05$.

\section*{Conclusions}
 In this paper we have introduced the concept and quantitative measure of fan-out for genetic circuits which is a measure of the maximum number of promoter sites that  the output TFs  of the upstream module can regulate without significant slow-down in the kinetics of the output. In addition we have also proposed an efficient method to estimate the fan-out experimentally. We have shown that the fan-out can be enhanced by self-inhibitory regulation on the output. In the estimation process of the fanout, the retroactivity can also be estimated.    This study provides a way for quantifying the level of modularity in gene regulatory circuits and helps characterize and design module interfaces and therefore the modular construction of gene circuits.
  
\section*{Methods}
  We obtain the time-constant $\tau_{P_T}$ and derive the fan-out functions for various cases.

\subsection*{Monomer TF } \noindent
Consider the simple MIP shown in Fig.~3.  Under the quasi-equilibrium in the binding-unbinding process, we obtain the concentration of the bound TF as $P_b = P_T X/ (X+K_d)$. We will simplify this equation to, by introducing $f(X) \equiv X/(X+K_d)$,
\begin{equation}
P_b = P_T f(X).
\label{eqn:Pb}
\end{equation}
The time evolution of the total transcription factor ($Y= X+P_b$) is governed by the following equation \cite{Kim2009}:
\[
\frac{dY}{dt} = \alpha - \gamma X.
\]
We will obtain the response time constant of $Y$ rather than $X$, because $Y$ is a pure slow variable showing the dynamics of our interest \cite{Rao2003, Vecchio2008, Kim2009}.  We denote the response time constant of $Y$ by $\tau_{P_T}$, and is obtained by taking the derivative on the right hand side of the above equation with respect to $Y$:
\begin{equation}
\tau_{P_T} = -\left[ \frac{d(\alpha - \gamma X)}{dY}   \right]^{-1} =  \tau_0 \frac{dY}{dX},
\label{eqn:Pt}
\end{equation}
with $\tau_0 \equiv \gamma^{-1}$.  By using $Y=X+P_b$, the above equation becomes $\tau_{P_T} = \tau_0 \left[ 1+ \frac{dP_b}{dX}\right]$.
By using Eq.~\eqref{eqn:Pb}, we obtain the time constant: $\tau_{P_T} = \tau_0 ( 1+ f'(X) P_T)$. By comparing this with Eq.~\eqref{eqn:response-time} we obtain $C/C_1 = K_d (1+X/K_d)^2$.

\subsection*{Oligomer TF under directed degradation and self-regulation}\noindent
We consider that the transcription factors are composed of $n$ monomers described in Fig.~5A.  We assume that the binding-unbinding process is in equilibrium, and we obtain $P_b$ as 
\[
P_b= \frac{P_T X^n}{X^n+K_d} \equiv P_T f(X).
\] 
The time evolution of $Y(= X+ nP_b)$ is governed by 
\[
\frac{dY}{dt} = v_1(X) - v_2(X).
\]
 The response time constant of $Y$ is obtained as 
\[
\tau_{P_T} = -\left[ \frac{d[v_1(X) - v_2(X)]}{dY}   \right]^{-1} = \tau_0 \frac{dY}{dX},
\]
where $\tau_0 \equiv 1/(\epsilon_2-\epsilon_1)$ with $\epsilon_1 \equiv \partial v_1/ \partial X$ and $\epsilon_2 \equiv d v_2(X)/ d X$.  By using $Y=X+nP_b$, we obtain 
\[
\tau_{P_T} = \tau_0 \left[   1+  P_T ~n \frac{df(X)}{dX}\right],
\]
where $n  \frac{df(X)}{dX}$ is defined to be $C_1/C$ by comparing the above equation with Eq.~\eqref{eqn:response-time}.

\subsection*{Multiple promoters having different affinities }\noindent
We consider the case that two different types of TF-specific promoter plasmids, having different affinities for the TF and different strength of the origin of replication.  The TF is assumed to be a monomer.   The concentration of the TF bound on the promoter of each type ($i$) is given as $P_{bi} = \frac{P_{Ti} X}{X+K_{di}} \equiv P_{Ti} f_i(X)$ 
for $i=1,2$.  The response time constant $\tau_{P_{T1}, P_{T2}}$ is given by Eq.~\eqref{eqn:Pt}.  By using $Y = X+P_{b1}+ P_{b2}$, we obtain $\tau_{P_{T1}, P_{T2}} = \tau_0 \left[  1+ \frac{dP_{b1}}{dX} + \frac{dP_{b2}}{dX}\right]$,
and rewrite this as $\tau_{P_{T1}, P_{T2}} = \tau_0 \left[  1+ f_{1}'(X) P_{T1}+ f_{2}'(X) P_{T2}\right]$ by using $P_{bi} =  P_{Ti} f_i(X)$.
Finally by equating $\tau_{P_{T1}, P_{T2}}$ to $1/\omega_{max}$, we obtain Eq.~\eqref{eqn:fan-out-gen}.  If there were $n$-different types of promoter plasmids, the above equation is changed by replacing the last two terms to the sum over all the $n$-types. 

In the RC-circuit representation, $\tau_{P_{T1}, P_{T2}}$ is given by $RC_{T}$ and the total capacitance $C_{T}$ is obtained as $C_T =  C + C_1 P_{T1} + C_2 P_{T2}$
with $C_i = \tau_0 f_i'(X)/R$.  This indicates that this MIP can be mapped to an RC-circuit having two different capacitances connected in parallel to $C$ as shown in Fig.~5B.

\section*{Competing interests}\noindent
The authors declare that they have no competing interests.

\section*{Authors contributions}\noindent
KHK primarily developed and carried out the proposed work.  HS participated its design and provided critical feedback and suggestions. All authors read and approved the final manuscript.

\section*{Acknowledgments}\noindent
This work was supported by a National Science Foundation (NSF) Grant in Theoretical Biology 0827592. Preliminary studies were supported by funds from NSF FIBR 0527023.  The authors acknowledge useful discussions with Hong Qian and Suk-jin Yoon.



\pagebreak




\section*{Supplementary Information}\noindent
\section{Repressilator BioModel}	
We obtained a model for the repressilator from the BioModels Database  (model id: BIOMD0000000012) \cite{Novere2006}.  This model is composed of transcription and translation processes for respective \emph{lacI}, \emph{tetR}, and \emph{cI}.  Inhibitory regulations among them are described by Hill functions.  One of the repressors, TetR, is chosen as an output of the repressilator module.  The TetR repressors are allowed to regulate a downstream module and its regulation is described by the binding-unbinding reactions between the TetR and its specific promoter located in the downstream module.  The copy numbers of mRNAs of respective \emph{lacI}, \emph{tetR}, and \emph{cI} are denoted by $X$, $Y$, and $Z$, and the corresponding  repressor molecules by $PX$, $PY$ and $PZ$.   The model process (per cell) is described as, 
\begin{eqnarray*}
&&\xrightarrow{a_{0tr} + a_{tr}\frac{K_M^2}{K_M^2+PZ^2}} X \xrightarrow{kd_{mRNA}X} \o,\\
&&\xrightarrow{a_{0tr} + a_{tr}\frac{K_M^2}{K_M^2+PX^2}} Y \xrightarrow{kd_{mRNA}Y} \o,\\
&&\xrightarrow{a_{0tr} + a_{tr}\frac{K_M^2}{K_M^2+PY^2}} Z \xrightarrow{kd_{mRNA}Z} \o,\\
&&\xrightarrow{k_{tl}X} PX \xrightarrow{kd_{prot}PX} \o,\\
&&\xrightarrow{k_{tl}Y} PY \xrightarrow{kd_{prot}PY} \o,\\
&&\xrightarrow{k_{tl}Z} PZ \xrightarrow{kd_{prot}PZ} \o,\\
&&2 PY + P_f \xrightleftharpoons[k_{off} P_b]{k_{on}PY^2 P_f} P_b,
\end{eqnarray*}
where the total number $P_T$ of the \emph{tetR} promoter in the downstream module  is given by the sum of the numbers of the free and bound promoters: $P_f + P_b$.  We have assumed that the transcription and translation processes for each repressor gene are identical.  
The parameter values used for simulations are listed in Table \ref{table:param}.  We have modified the parameters to increase the retroactivity by reducing the expression levels of the repressor molecules ($a_{tr}: 29.97 \rightarrow 2.97$, $K_M: 40 \rightarrow 10$, $eff: 20 \rightarrow 10$).  

\begin{table*}[t]
\centering
\begin{tabular}{@{\vrule height 10.5pt depth4pt  width0pt}lll}
\hline \hline
$a_{0tr}$ & $0.03$ min$^{-1}$ & Basal transcription rate\\
$a_{tr}$ &  $3-0.03=2.97$ min$^{-1}$  & Maximum transcription rate is set to 3 min$^{-1}$.\\
$K_M$    &  $10$ & Number of  repressor molecules corresponding to\\
&&  the half maximal transcription\\
$kd_{mRNA}$ & $\frac{\log(2)}{\tau_{mRNA}}$ min$^{-1}$& Degradation rate constant of mRNA\\
$\tau_{mRNA}$ & 2 min & Half life time of mRNA \\
$k_{tl}$ & $eff \frac{\log(2)}{\tau_{mRNA}}$ min$^{-1}$ & Translation rate per mRNA\\
$eff$ & 10 & Translation efficiency: Number of repressor molecules \\
&&translated per mRNA during mRNA life time\\
$kd_{prot}$ & $\frac{\log(2)}{\tau_{prot}}$  min$^{-1}$ & Degradation rate constant of repressor molecules \\
$\tau_{prot}$ & 10 min & Repressor molecule half life time\\
$k_{on}$ & $0.166$ min$^{-1}$ & Promoter-repressor binding constant \cite{Elf2007}\\
$k_{off}$ &  $1.66$ min$^{-1}$ & Promoter-repressor unbinding constant\\
&& for a given dissociation constant $K_d = k_{off}/k_{on} = 10$\\
$P_T$ & 0  or 100  & Number of \emph{tetR} promoters in a downstream module\\
\hline 
\hline
\end{tabular}
\caption{Parameters of the repressilator BioModel}
\label{table:param}
\end{table*}

\section{Linear relationship between $C'$ and $P_T$}
In this section, we consider the case that the \emph{intermediate} reaction steps of oligomerization and enzyme-mediated degradation are taken into account.
\subsection{Oligomer TF}\noindent
We consider a dimer TF (the results obtained below  can be generalized to any other oligomer types).  The corresponding model can be expressed as:
\begin{eqnarray*}
&\xrightarrow{~~\alpha ~~}X \xrightarrow{\gamma X} \o \\
&X+X  \xrightleftharpoons[k_2X_2]{k_1X^2}X_2 \xrightarrow{\gamma_2 X_2} \o \\
&X_2 + P_f  \xrightleftharpoons{K_d} P_{b},
\end{eqnarray*}
where the reactions from the top represent the monomer translation and degradation, dimerization, dimer degradation, and promoter-binding-unbinding reactions.  $K_d$ denotes the dissociation constant.  The time evolution of the total transcription factor (monomer units) ($Y= X+2X_2+2P_b$) is governed by 
\[
\frac{dY}{dt} = \alpha - \gamma X - 2 \gamma_2 X_2.
\]
The response time constant of $Y$ becomes
\begin{equation}
\tau_{P_T} = - \left[  \frac{d(\alpha - \gamma X - 2 \gamma_2 X_2}{dY} \right]^{-1} = \left[\gamma \frac{dX}{dY} + 2 \gamma_2 \frac{dX_2}{dY}\right]^{-1}.
\label{eqn:pt-dimer}
\end{equation}
Under the assumption of the equilibrium of $X_2$ and $P_b$, we obtain 
\begin{equation}
X_2 = \frac{k_1}{k_2 + \gamma_2} X^2,~~\mbox{and}~~ P_b = \frac{P_T X_2}{K_d + X_2}\equiv P_T f(X).
\label{eqn:x2}
\end{equation}
By substituting the first equation in the above to Eq.~\eqref{eqn:pt-dimer}, we obtain
\[
\tau_{P_T} = \left[ \gamma + \frac{4 \gamma_2 k_1 X} {k_2+ \gamma_2}\right]^{-1} \frac{dY}{dX}
\]
By using $Y= X+ 2X_2 + 2 P_b$ and Eq.~\eqref{eqn:x2}, we obtain
\begin{equation}
\tau_{P_T} = \tau_0 \left[  1+ \frac{2f'(X)}{1+\frac{4k_1 X}{k_2 + \gamma_2}}P_T  \right],
\label{eqn:taupt-oligo}
\end{equation}
where $\tau_0$ denotes the time constant in the case without any promoter:
\[
\tau_0 = \frac{1+\frac{4k_1 X}{k_2 + \gamma_2}}{\gamma + \frac{4 \gamma_2 k_1 X }{k_2 + \gamma_2} }.
\]
We note that  the steady state concentration of $X$ is independent of $P_T$.  Equation \eqref{eqn:taupt-oligo} shows that the extra capacitance $C'$ is proportional to $P_T$ (by noting that $\tau_{P_T} = RC_T = R(C+C')$ and $\tau_0 = RC$).

\subsection{Degradation-tagged TF}\noindent
We consider a monomer TF under directed degradation by proteases. The corresponding model can be expressed as:
\begin{eqnarray*}
&\xrightarrow{~~\alpha ~~}X \\
&X+ E \xrightleftharpoons[k_2 {[}XE{]}]{k_1 X \cdot E}XE \xrightarrow {\gamma [XE]} E \\
&X + P_f  \xrightleftharpoons{K_d} P_{b},
\end{eqnarray*}
with the total number of the proteases $E_T(=E+[XE])$ fixed.  The time evolution of the total TF ($Y=X+[XE]+P_b$) is governed by 
\[
\frac{dY}{dt} = \alpha - \gamma [XE].
\]
The response time constant of $Y$ becomes 
\[
\tau_{P_T} = - \left[  \frac{d(\alpha - \gamma [XE])}{dY}  \right]^{-1} = \left[  \gamma \frac{d [XE]}{dY} \right]^{-1}.
\]
Under the assumption of the equilibrium of $[XE]$ and $P_b$, we can express $[XE]$ and $P_b$ in terms of $X$ as like Eq.~\eqref{eqn:x2}.  By following the same way as above, we can prove that the time constant increases by the amount proportional to $P_T$ as like Eq.~\eqref{eqn:taupt-oligo}.

\subsection{Multiple operators }\noindent 
Consider a dimer TF that can bind two operator sites with different affinities. The binding-unbinding process between the TF and the promoter can be modeled as 
\begin{eqnarray*}
&X+P_0 \xrightleftharpoons{} P_1\\
&X+P_0  \xrightleftharpoons{} P_2\\
&P_1  \xrightleftharpoons{}P_{12}\\
&P_2  \xrightleftharpoons{} P_{12},
\end{eqnarray*}
where $P_0$, $P_i$, and $P_{12}$ denote promoters that are free, occupied at the $i$-th operator site, and fully occupied, respectively.  The equilibrium  constant of each reaction from top to bottom is denoted by $K_1$, $K_2$, $K_3$, and $K_{4}$.  By using that the total promoter concentration $P_T$ equals $P_0+ P_1 + P_2+P_{12}$, we obtain 
\begin{equation}
P_b \equiv P_1+ P_2 + P_{12} = \frac{P_T X}{X+K_d} \equiv P_T f(X),
\label{eqn:P12}
\end{equation}
where 
\[
K_d \equiv \left[  K_1+K_2+K_1/K_3  \right]^{-1}.
\]
We note that $K_1/K_3 = K_2/K_4$.  The response time constant $\tau_{P_T}$, given by
\[
\tau_{P_T} = -\left[ \frac{d(\alpha - \gamma X)}{dY}   \right]^{-1} =  \tau_0 \frac{dY}{dX},
\]
 is re-written as  
\begin{equation}
\tau_{P_T} = \tau_0 ( 1+ f'(X) P_T).
\label{eqn:Pt-app}
\end{equation}
 by using $Y=X+P_b$.  If there were $n$-operators and the TF is an oligomer, the same derivation can be applied resulting in  Eq.~\eqref{eqn:Pt-app}.

\bibliographystyle{bmc_article}
\bibliography{./research}

\begin{thebibliography}{10}
\bibitem{note-retroact}
 [The solution of Eq.~\eqref{eqn:TF-isolation} with $\alpha$ constant is given
  by $X(t)=\left( X(0) - \alpha/\gamma \right) e^{-\gamma t} + \alpha/\gamma$,
  with $X(0)$ the initial concentration. Thus, the exponential function
  decreases sufficiently at $t \simeq 1/\gamma$.].
\end{thebibliography}

\section*{Figure 1 -- Gene circuit modules and their interface process}
(A) The repressilator (Module 1) \cite{Elowitz2000} regulates multiple copies of a downstream module (Module 2).  The multiple copies can be realized by placing the downstream module in a plasmid.  TetR repressors (output of Module 1)  regulate their specific downstream-module promoters (input of Module 2).  A module interface process (MIP) defines the collection of  the processes of \emph{tetR}-transcription and translation and its specific downstream-module regulation.  As the number of the downstream-module promoter ($P_T$) increases, the amplitude and period of the oscillation in the repressilator can be changed (B and C).  A repressilator model was obtained from the BioModels Database BIOMD0000000012 \cite{Novere2006}. We modified the model to lower the expression levels (by changing translation efficiency to 10 and $K_M$ to 10  and the maximum transcription rate to 3 per min per cell) and to add promoter binding-unbinding reactions for TetR repressors (for the detailed model description, refer to the SI).

\section*{Figure 2 -- Isolated module output}
Translation-degradation processes for $X$ (A) can be described by a simple reaction process (B) with $\alpha$ a translation rate and $\gamma X$ a degradation rate.  These processes can be mapped to an RC-circuit with $R$ resistance and $C$ capacitance by $V_{out}=X$, $V_{in}=\alpha/\gamma$, and $RC=1/\gamma$ (C).  

\section*{Figure 3 -- Module interface process}
 The output $X$ of an upstream module regulates the downstream-module $X$-specific promoter (A).  The translation-degradation processes for $X$ and its promoter regulation  can be modeled as the  reaction process shown in B, where $P_f$, $P_b$, and $P_T$ denote the numbers of free, bound, and total promoters, respectively.  The reaction process is mapped to an RC-circuit with an increased capacitance by $C'$, which is shown to be proportional to $P_T$: $C'=P_T C_1$ with $C_1$ a proportionality constant.  This means that each promoter acts as a capacitor with a unit load of capacitance, $C_1$.  The total capacitance $C_T$ becomes the sum of the capacitance $C$ in the isolated  case and the extra capacitance $C'$.

\section*{Figure 4 -- Frequency response of the module interface process shown in Fig.~3A}
An oscillatory signal is applied at $V_{in}$ with different frequencies (A).  The signal gain is defined by the ratio of the oscillation amplitude of the output signal ($V_{out}$) to that of the input ($V_{in}$): $g(\omega) = \frac{\Delta V_{out}(\omega)}{\Delta V_{in}(\omega)}$, and    can be well approximated by  $g(\omega) = \sqrt{1+\omega^2/(RC_T)^2}^{-1}$  \cite{Nilsson2008} with $C_T$ the total capacitance.  As the number of the promoters that the output (TF) signal drives (regulates) increases, the cut-off frequency ($\omega_c = 1/RC_T$) decreases (B).  We assume that the output signal is desired to be operated within a certain frequency range between 0 and 1 hour$^{-1}$, i.e., the maximum operating frequency $\omega_{max}$ is 1 hour$^{-1}$.  When the cut-off frequency matches the maximum operating frequency, the corresponding number of the promoters is defined as the fan-out (C).  Parameters of the model: $K_d=1$ nM [$k_{on}$ = $10  (1/\mbox{nM/hour})$, $k_{off} = 10$ $ (1/\mbox{hour})$],  $\gamma = 2 (1/\mbox{hour})$, $\alpha=20 (\mbox{nM}/\mbox{hour})$.

\section*{Figure 5 -- Module interface processes that the fan-out function Eqs.~\eqref{eqn:fan-out} and \eqref{eqn:fan-out-gen} can be applied to}
(A) An oligomer TF degraded by proteases. (B) A TF can bind two different promoter plasmids having different binding affinities and different origins of replication. This can be mapped to an RC-circuit with two different capacitances connected in parallel. (C) An Oligomer TF can bind multiple operators. (D) Each different TF binds to its specific operator without affecting the binding affinity of the other.

\begin{figure}[t!]
  \centering
  \includegraphics[width=5in]{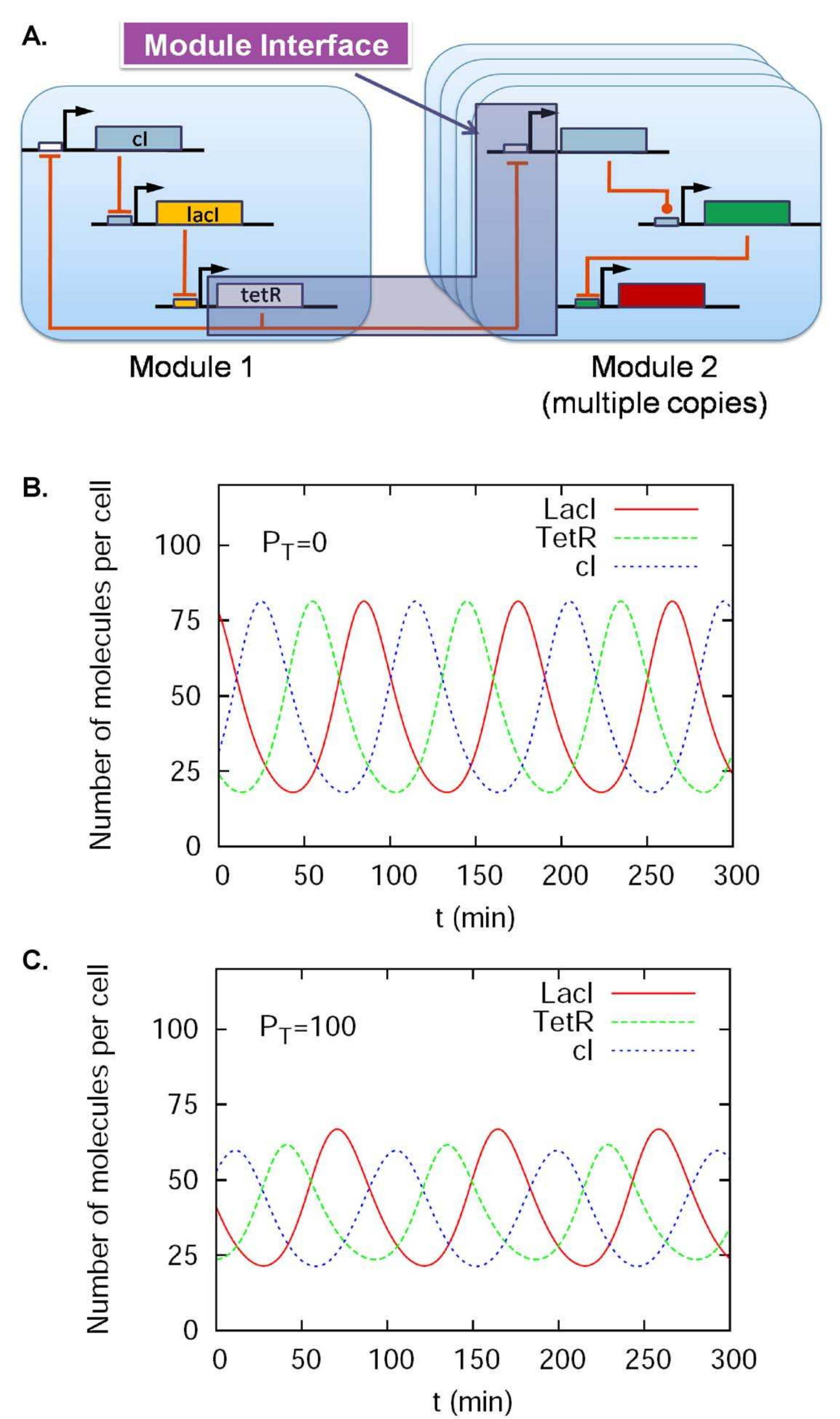}
  \caption{} \label{fig:module}
\label{fig:module}
\end{figure}

\begin{figure}[t]
  \centering
 \includegraphics[width=5in]{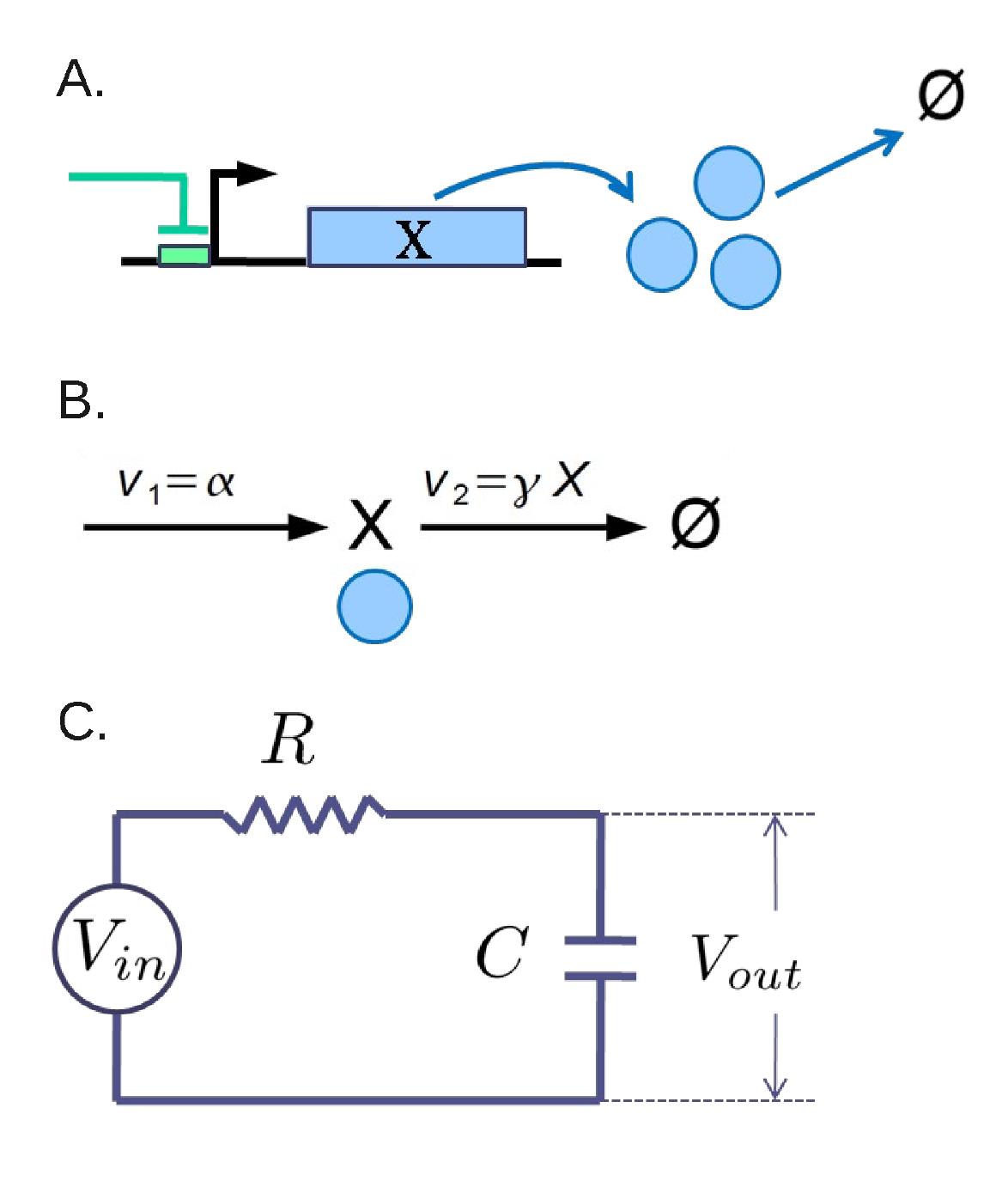}
  \caption{ } 
\label{fig:IsolatedModule}
\end{figure}

\begin{figure*}[t]
  \centering
  \includegraphics[angle =90, width=5in]{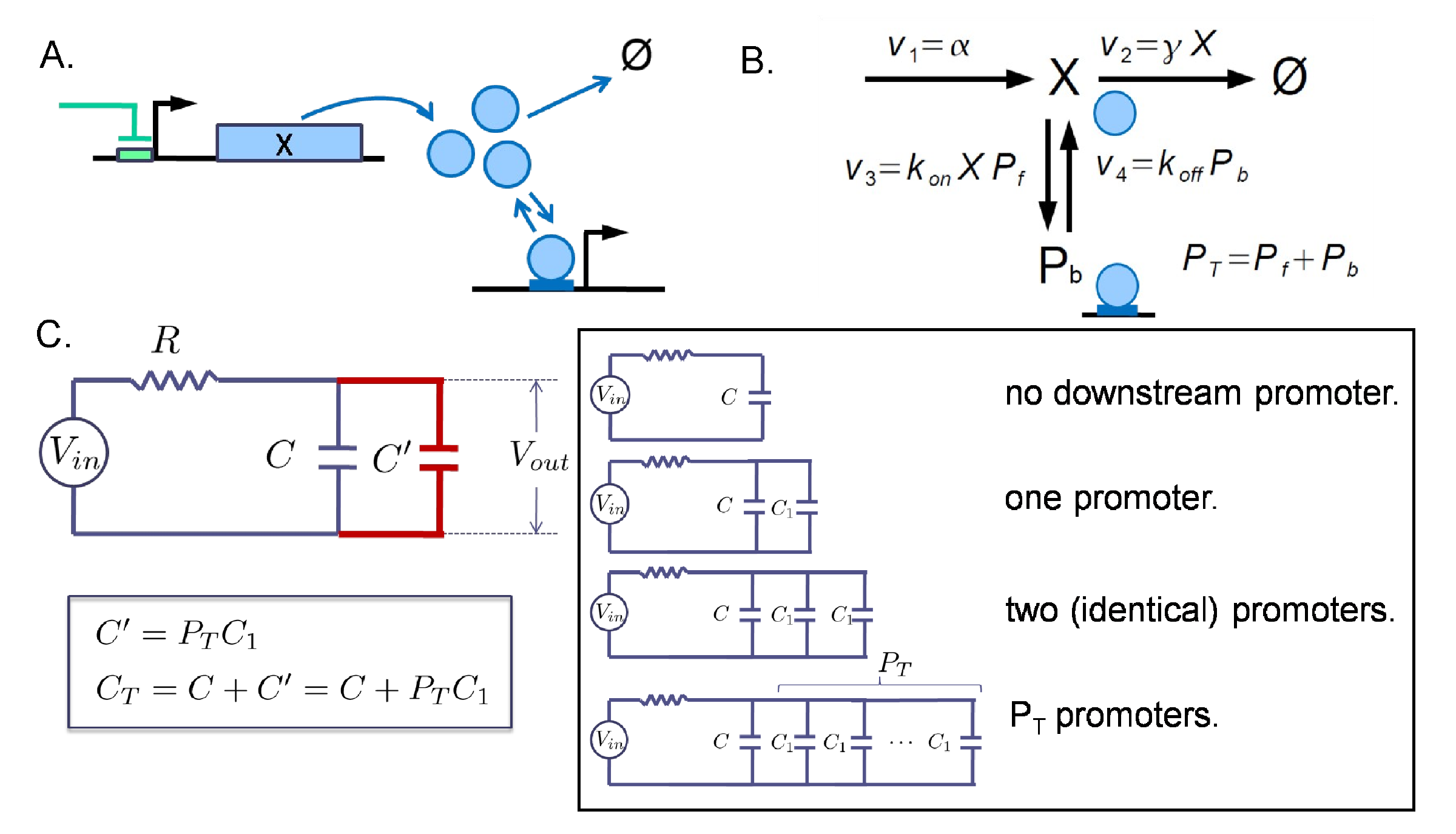}
  \caption{} 
\label{fig:ConnectedModule}
\end{figure*}

\begin{figure*}[t]
  \centering
  \includegraphics[width=7in]{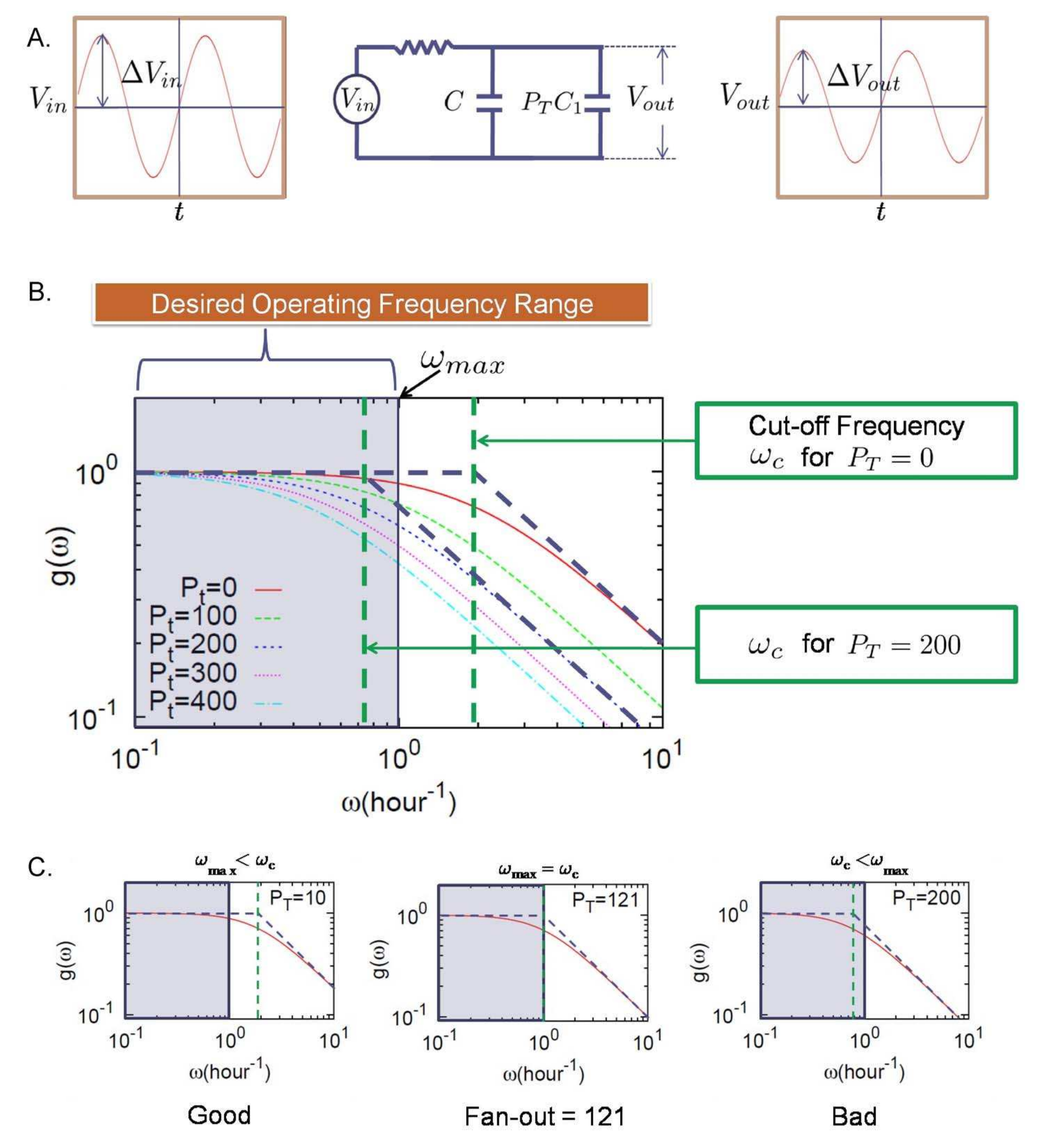}
  \caption{ }
 \label{fig:FrequencyResponse}
\end{figure*}

\begin{figure*}[ht!]
  \centering
  \includegraphics[width=7in]{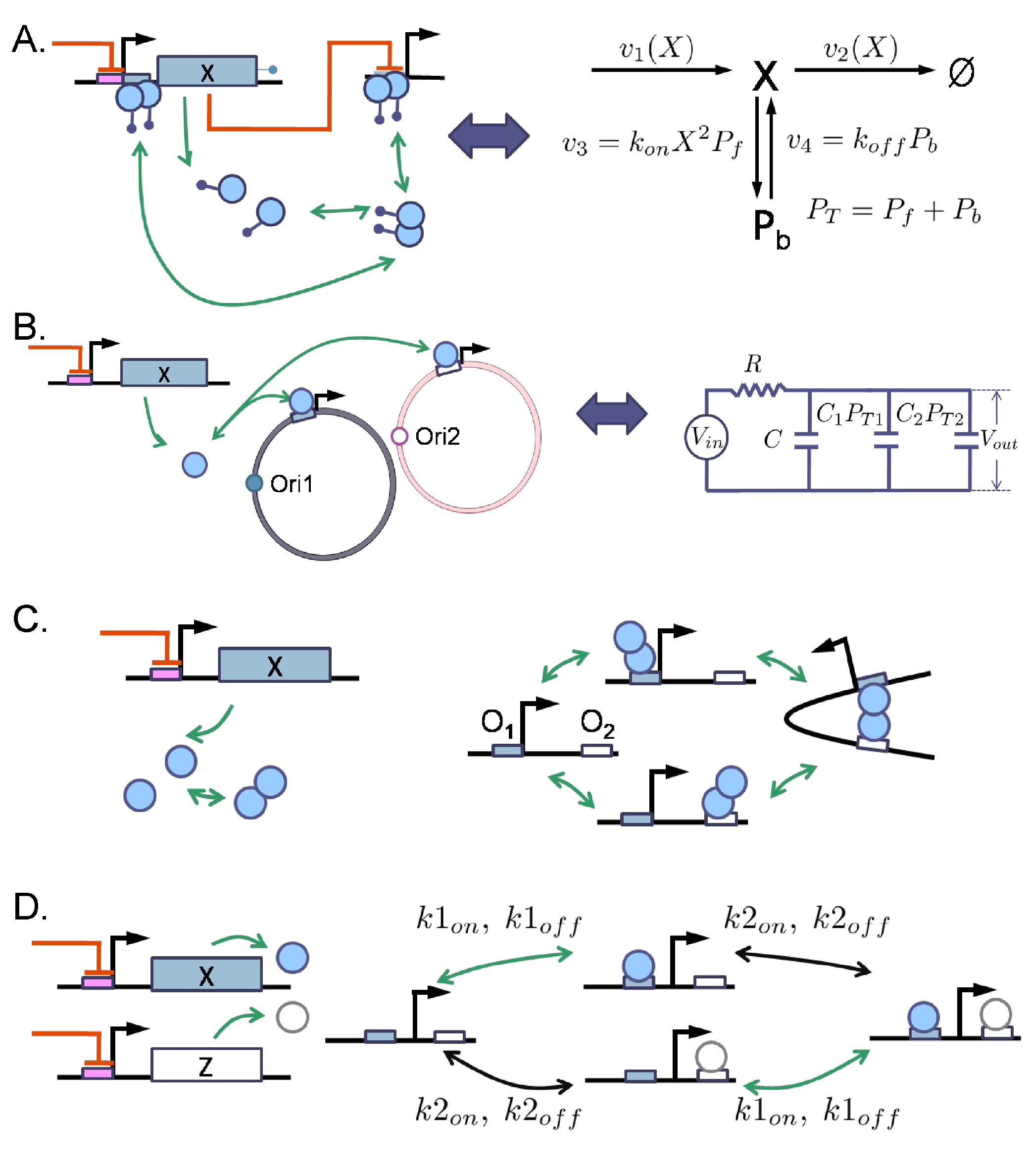}
  \caption{} 
\label{fig:ConnectedModuleGen}
\end{figure*}

\end{document}